# TIME ENCODING SAMPLING OF BANDPASS SIGNALS

*Zhong Liu, Feng Xi, and Shengyao Chen*

Department of Electronic Engineering
Nanjing University of Science and Technology
Nanjing, Jiangsu 210094, PRC

**ABSTRACT**

This paper investigates the problem of sampling and reconstructing bandpass signals using time encoding machine (TEM). It is shown that the sampling in principle is equivalent to periodic non-uniform sampling (PNS). Then the TEM parameters can be set according to the signal bandwidth and amplitude instead of upper-edge frequency and amplitude as in the case of bandlimited/lowpass signals. For a bandpass signal of a single information band, it can be perfectly reconstructed if the TEM parameters are such that the difference between any consecutive values of the time sequence in each channel is bounded by the inverse of the signal bandwidth. A reconstruction method incorporating the interpolation functions of PNS is proposed. Numerical experiments validate the feasibility and effectiveness of the proposed TEM scheme.

## 1. INTRODUCTION

Time encoding machine (TEM) is a biological neuron-like paradigm to perform signal sampling. Different from the traditional sampling by measuring amplitudes of a continuous-time signal at pre-defined sampling times, TEM makes sampling by recording the time at which the signal or its function takes on a preset value. TEM was firstly proposed for the sampling and reconstruction of the bandlimited signals [1,2]. Its initial success was then extended to the cases of signals in shift-invariant subspaces [3], finite-rate-of-innovation signals [4,5], and even non-bandlimited signals [6]. Signal processing with time encoding sampling also attracts attention [7, 8].

TEMs are event-driven sampling schemes and have different structures, such as integrate-and-fire time encoding machine (IF-TEM) [1], crossing time encoding machine [3], differentiate-and-fire time encoding machine [9], and so on. In this paper, we are interested in IF-TEM described by three parameters: bias, scale and threshold [1]. In operation, TEM adds a bias to its input, scales the sum and integrates the result, then compares the integral value with a threshold. A threshold-crossing or spiking time is recorded when the integral reaches the threshold.

TEM outputs a time sequence consisting of strictly increasing times and the information of signal is encoded in the spiking sequence. Then a fundamental problem is if the sequence contains the information enough to reconstruct the signal from the sequence or under what conditions the signal can be reconstructed from the sequence. Interestingly, it is found that for the bandlimited signal, it can be perfectly reconstructed if the difference between any two consecutive values of the time sequence is bounded by the inverse of the Nyquist rate [1]. That is to say that the spiking interval between any two consecutive spiking times should be smaller than or equal to the inverse of the Nyquist rate. This condition is closely related to the traditional Shannon-Nyquist sampling theory [10]. For convenience, we call the *largest interval* as *TEM interval* in this paper.

TEM interval is equal to the inverse of the Nyquist rate of the signal. For large bandwidth signals, TEM with small TEM interval is needed for the signal reconstruction. This is much inconvenient for the selection of TEM parameters as in traditional sampling by which high-speed analog-to-digital converters are required. Then multi-channel TEM is developed [11~13]. It is found that for an *M*-channel TEM with shifted integrators, perfect reconstruction is possible with *M* times the TEM interval of the single channel case. This result is much like Papoulis's multi-channel sampling criterion in the traditional sampling setup [14].

In this paper, we study the time encoding of bandpass signals. Traditionally, the signals can be sampled by several schemes with the minimum Landau's rate as discussed in [15]. Among them, periodic nonuniform sampling (PNS) scheme [16] is a popular one due to its hardware simplicity. PNS consists of multiple sampling channels and outputs multiple sampling sequences with relative time-shifts. By properly setting the time-shifts and the interpolation functions, the bandpass signal can be reconstructed. What is important is that the sampling period in each channel is set according to the signal bandwidth and the channel numbers, and the minimal Landau's rate can be achieved [15]. Although the multi-channel TEM and the PNS scheme are formulated from different mechanisms, they share a common property of *shifts*. The shifts in PNS imply the time shifts between different sampling channels and thus gener-

This work was partially supported by Natural Science Foundation of China (62171224).

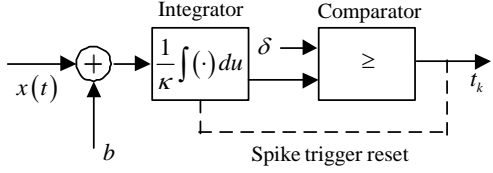

Fig.1. Time encoding machine with spike trigger reset

ate different undersampling sequences. In multi-channel TEM, the shifts refer to the integrator shifts between different TEM channels, which in turn introduce the time shifts in recording time instants and then generate different spiking sequences. It is this observation that makes us perform the time encoding sampling of the bandpass signals via multi-channel TEM with TEM interval determined by the signal bandwidth instead of the upper frequency. Then the largest TEM interval can be approached. We will take the bandpass signals consisting of a single information band as an example for the following discussions. The two-channel TEM in [12] is used to perform the TEM sampling and a reconstruction scheme incorporating the interpolation function in [17] is developed.

In the following of this paper, the fundamentals of TEM and PNS are firstly reviewed. Then a multi-channel TEM is discussed and a reconstruction algorithm is provided. Finally, simulation results are presented and conclusions are drawn.

## 2. PRELIMINARIES: TEM AND PNS

In this section, we will summarize main results of TEM and PNS from [1] and [16,17], respectively.

### 2.1 TEM

The TEM or single-channel TEM is shown in Fig.1. It consists of adder, integrator and comparator. Three parameters $\kappa$, $\delta$ and $b$ are used to describe the structure. For input signal $x(t)$, TEM adds a bias $b$ to it, scales the sum by $1/\kappa$ and then integrates the result until a threshold $\delta$. When this threshold is reached, a time is recorded, the value of the integrator is reset to $-\delta$. To keep regular operation, the signal $x(t)$ is assumed to be bounded, $|x(t)| \leq c$, and the bias $b$ is set to be $b > c$ so that the integrator output is a positive increasing function of time. In this way, TEM outputs a time sequence consisting of strictly increasing times $\{t_k | k \in \mathbb{Z}\}$ with $t_{k+1} > t_k$, and it satisfies

$$\frac{1}{\kappa} \int_{t_k}^{t_{k+1}} (x(u) + b) du = 2\delta \quad (1)$$

Defining $y_k \triangleq \int_{t_k}^{t_{k+1}} x(u) du$, we have

$$y_k = 2\kappa\delta - b(t_{k+1} - t_k) \quad (2)$$

In this way, we derive a amplitude-integral sequence of $\{y_k | k \in \mathbb{Z}\}$ from the time encoding $\{t_k | k \in \mathbb{Z}\}$ and TEM parameters $\{\kappa, \delta, b\}$. Because of non-uniform distances between any two consecutive spiking times, the sequence $\{y_k | k \in \mathbb{Z}\}$ resembles irregular sampling sequence of $x(t)$.

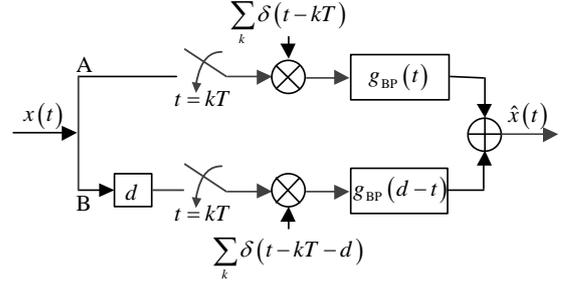

Fig.2 Two-channel periodic nonuniform sampling

In the TEM theory, $\{(t_k, y_k) | k \in \mathbb{Z}\}$ is used to reconstruct $x(t)$. Note that $|x(t)| \leq c$. Then from (2), we have

$$t_{k+1} - t_k \leq \frac{2\kappa\delta}{b-c} \quad (3)$$

That is, the difference between any two consecutive values of the sequence $\{t_k | k \in \mathbb{Z}\}$ less than or equal to $2\kappa\delta/(b-c)$. For convenience, we define $T_{\text{TEM}}^{\text{LP}} \triangleq 2\kappa\delta/(b-c)$ and name it as the *TEM interval*. For a $2\Omega$-bandlimited signal, i.e., its Fourier transform is zero for $|\omega| \in (\Omega, \infty)$, it has been proved [1] that perfect reconstruction is possible if

$$T_{\text{TEM}}^{\text{LP}} < \frac{1}{F_{\text{NYQ}}} \quad (4)$$

where $F_{\text{NYQ}} = \Omega/\pi$ is the Nyquist sampling rate of the signal $x(t)$. It is seen that for perfect reconstruction, the TEM interval is bounded by the inverse of the Nyquist rate. To be different from the bandpass signal discussed in this paper, we use the superscript 'LP' to denote it for the bandlimited/lowpass signals.

To reconstruct the signal $x(t)$, a possible way is to do as in irregular sampling. Let

$$x(t) = \sum_{\ell} c_\ell g_{\text{LP}}(t - s_\ell) \quad (5)$$

where $s_\ell = (t_\ell + t_{\ell+1})/2$, $g_{\text{LP}}(t) = \sin(\Omega t)/(\pi t)$ and $\{c_\ell\}$ are coefficients to be determined. From (5), we have

$$\int_{t_k}^{t_{k+1}} x(u) du = \int_{t_k}^{t_{k+1}} \sum_\ell c_\ell g_{\text{LP}}(u - s_\ell) du$$
$$= \sum_\ell c_\ell \int_{t_k}^{t_{k+1}} g_{\text{LP}}(u - s_\ell) du$$

Then by (2), we get

$$\sum_\ell c_\ell \int_{t_k}^{t_{k+1}} g_{\text{LP}}(u - s_\ell) du = y_k$$

from which a matrix equation $\mathbf{Gc} = \mathbf{q}$ can be established,

$$\mathbf{q} = [y_k]_{k \in \mathbb{Z}} \text{ and } \mathbf{G} = [G_{k\ell}]_{k,\ell \in \mathbb{Z}} = \left[\int_{t_k}^{t_{k+1}} g_{\text{LP}}(u - s_\ell) du\right]_{k,\ell \in \mathbb{Z}}$$

Then the coefficients $\{c_\ell\}$ can be estimated as $\mathbf{c} = [c_\ell]_{\ell \in \mathbb{Z}} = \mathbf{G}^+ \mathbf{q}$ where $\mathbf{G}^+$ is the pseudo-inverse of $\mathbf{G}$.

### 2.2 PNS

We consider a bandpass signal $x(t)$ with its support on $(\omega_l, \omega_u) \cup (-\omega_u, -\omega_l)$ and $B = \omega_u - \omega_l$ as its bandwidth.

PNS with two channels is shown in Fig.2, where the channels A and B assumes periodic sampling with sampling period $T$ and $d \in (0,T)$ is the time-shift. In the following, we assume that $T = 2\pi/B$ and then the Landau' rate is approached.

From Fig.2, it is seen that the sampling instants of A are always $d$ ahead of B. Note that the sampling period is larger than the inverse of Nyquist rate. Then aliases of the band contents are inevitable and it is impossible to perfectly reconstruct the $x(t)$ by directly combining the two channel samples. However, it is shown that perfect reconstruction is possible if $d$ is such that $dK_0/T$ is not an integer where $K_0 = \lceil 2\omega_l/B \rceil$, and the signal $x(t)$ can be reconstructed by

$$x(t) = \sum_\ell x_\ell g_{BP,\ell}(t - \ell T, d) \qquad (6)$$

where $x_\ell$ is formulated from $x(kT)$ and $x(kT+d)$ as

$$x_\ell = \begin{cases} x(kT), & \ell = 2k \\ x(kT+d), & \ell = 2k+1 \end{cases} \qquad (7)$$

and $g_{BP,\ell}(t,d)$ is defined as

$$g_{BP,\ell}(t,d) = \begin{cases} g_{BP}(t,d), & \ell = 2k \\ g_{BP}(d-t,d), & \ell = 2k+1 \end{cases} \qquad (8)$$

with

$$g_{BP}(t,d) = \frac{\cos\left[(\omega_l + B)t - (K_0 + 1)Bd/2\right]}{Bt \sin\left[(K_0 + 1)Bd/2\right]} - \frac{\cos\left[(K_0 B - \omega_l)t - (K+1)Bd/2\right]}{Bt \sin\left[(K+1)Bd/2\right]} + \frac{\cos\left[(K_0 B - \omega_l)t - K_0 Bd/2\right]}{Bt \sin\left[K_0 Bd/2\right]} - \frac{\cos\left[\omega_l t - K_0 Bd/2\right]}{Bt \sin\left[K_0 Bd/2\right]} \qquad (9)$$

Note that we take a representation of (6) which is in the form different from that in [17].

PNS is an alternative sampling framework among multiple channel signals. In fact, we can combine and order the sampling instants as

$$t_\ell = \begin{cases} kT, & \ell = 2k \\ kT+d, & \ell = 2k+1 \end{cases} \qquad (10)$$

Then we have irregular samples $x(t_\ell)$. Ref. [18] studies the irregular sampling from the PNS theory and guarantees the reconstruction of $x(t)$ when $t_{\ell+1} - t_\ell < T$.

## 3. BANDPASS TEM: SAMPLING AND PERFECT RECONSTRUCTION OF BANDPASS SIGNALS

We consider a two-channel TEM in Fig.3, in which TEM A and TEM B are set to have the same TEM parameters but with different integrators. As discussed in [12], the thresholds of the two TEMs will be reached at different times. Then A and B will record different spiking times and produce different encoding of $x(t)$. In particular, the two spiking sequences will be interleaved in time, i.e., there is always one spike time from B between any two spikes of A, and vice versa. The interleaved sequences are inputted to reconstruction system to reconstruct the $x(t)$.

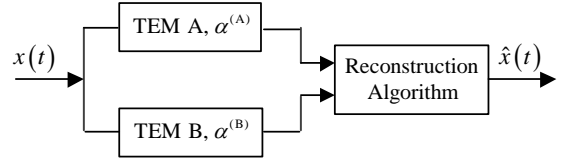

Fig.3 Two-channel time encoding and decoding

Assume that the integrator of A is $\alpha^{(A)} \neq 0$ ahead of the integrator B (modulo $2\delta$). It follows that the integrator of B is $\alpha^{(B)} = 2\delta - \alpha^{(A)}$ ahead of the integrator A. For convenience, we let $\delta < \alpha^{(A)} \leq 2\delta$. Then from the same starting time, the TEM A will output a spiking time ahead of TEM B. Denote the spike times of A and B as $\{t_k^{(A)}, k \in \mathbb{Z}\}$ and $\{t_k^{(B)}, k \in \mathbb{Z}\}$, respectively. We have $t_k^{(A)} < t_k^{(B)}, \forall k \in \mathbb{Z}$. The amplitude-integral sequences of A and B are given by

$$y_k^{(A)} = \int_{t_k^{(A)}}^{t_{k+1}^{(A)}} x(u)du = 2\kappa\delta - b\left(t_{k+1}^{(A)} - t_k^{(A)}\right)$$
$$y_k^{(B)} = \int_{t_k^{(B)}}^{t_{k+1}^{(B)}} x(u)du = 2\kappa\delta - b\left(t_{k+1}^{(B)} - t_k^{(B)}\right) \qquad (11)$$

We now combine and order the spike times $t_k^{(A)}$ and $t_k^{(B)}$ into one set of spike times $\{\tilde{t}_\ell, \ell \in \mathbb{Z}\}$,

$$\tilde{t}_\ell = \begin{cases} t_k^{(A)}, & \ell = 2k \\ t_k^{(B)}, & \ell = 2k+1 \end{cases} \qquad (12)$$

Then we can formulate an alternative set of discrete representations $\{\tilde{y}_\ell, \ell \in \mathbb{Z}\}$ defined as

$$\tilde{y}_\ell = \int_{\tilde{t}_\ell}^{\tilde{t}_{\ell+2}} x(u)du = \begin{cases} y_k^{(A)}, & \ell = 2k \\ y_k^{(B)}, & \ell = 2k+1 \end{cases} \qquad (13)$$

By comparing the $\tilde{y}_\ell$ with the PNS-generated irregular samples $x(t_\ell)$, we may find that there are strong similarities between them. Then we can argue that the signal $x(t)$ can be reconstructed from the sequence generated by two-channel TEM when $\tilde{t}_{\ell+1} - \tilde{t}_\ell < T$. If we define the TEM interval as $T_{TEM}^{BP} \triangleq 2\kappa\delta/(b-c)$ for the two-channel TAM, the signal $x(t)$ can be reconstructed if

$$T_{TEM}^{BP} < \frac{2\pi}{B} \qquad (14)$$

The TEM parameters can be designed according to the signal bandwidth instead of its upper-edge frequency.

Although the multi-channel TEM and the PNS scheme are formulated from different mechanisms, they share a common property of *shift*. As discussed in last section, the shift in PNS introduces the irregular sampling instants and thus generates irregular samples. For multi-channel TEM, the integrator shifts introduce the time shifts in recording time points and then generate the irregular sequence (13). The difference is that the time shift in PNS is fixed while the shift is varying from one spiking time to next one.

For the reconstruction, we consider the representation equation (6). First define two sequences $\tilde{s}_\ell = (\tilde{t}_\ell + \tilde{t}_{\ell+2})/2$ for $\ell \in \mathbb{Z}$ and $\tilde{d}_\ell = \tilde{d}_{\ell+1} = \tilde{s}_{\ell+1} - \tilde{s}_\ell$ for odd indexes $\ell$. The sequence $\{\tilde{s}_\ell\}$ is the same as that by combining and order-ing $s_k^{(A)} = \left(t_{k+1}^{(A)} - t_k^{(A)}\right)/2$ and $s_k^{(B)} = \left(t_{k+1}^{(B)} - t_k^{(B)}\right)/2$. The sequence $\{\tilde{d}_\ell\}$ represents the time delay of TEM B behind TEM A between spiking times $k$ and $k+1$. Then from two sequences defined by (12) and (13), as in (5), we have the representation of $x(t)$ as

$$x(t) = \sum_\ell \tilde{c}_\ell g_{BP,\ell}\left(t - \tilde{s}_\ell, \tilde{d}_\ell\right) \quad (15)$$

where $\{\tilde{c}_\ell\}$ are coefficients to be determined and $g_{BP,\ell}(t,d)$ is defined as in (8). Following the same line as in last section, we have

$$\int_{\tilde{t}_\ell}^{\tilde{t}_{\ell+2}} x(u) du = \int_{\tilde{t}_\ell}^{\tilde{t}_{\ell+2}} \sum_\ell \tilde{c}_\ell g_{BP,\ell}\left(u - \tilde{s}_\ell, \tilde{d}_\ell\right) du$$
$$= \sum_\ell \tilde{c}_\ell \int_{\tilde{t}_\ell}^{\tilde{t}_{\ell+2}} g_{BP,\ell}\left(u - \tilde{s}_\ell, \tilde{d}_\ell\right) du$$

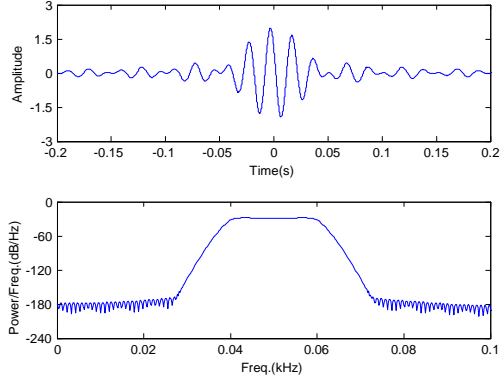

Fig.4 Signal and its power spectral density

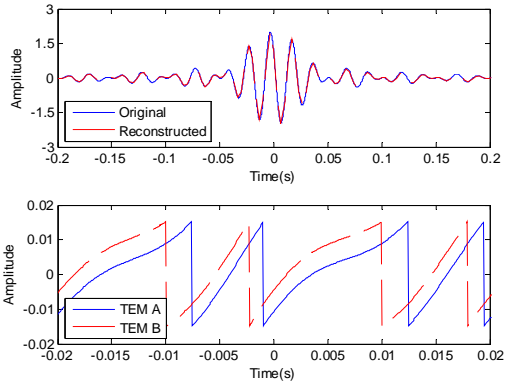

Fig.5 Two-Channel TEM: Signal and its reconstruction (upper) and integrator outputs (lower)

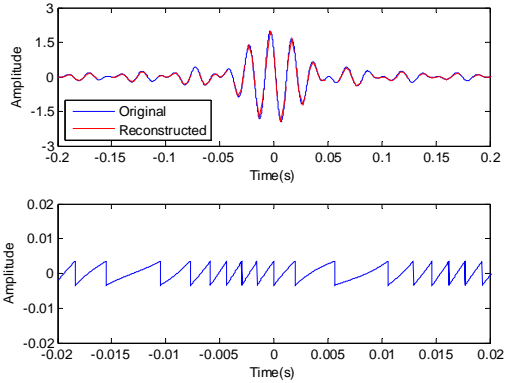

Fig.6 Single-Channel TEM: Signal and its reconstruction (upper) and integrator output (lower)

Then by (13), we get

$$\sum_\ell \tilde{c}_\ell \int_{\tilde{t}_\ell}^{\tilde{t}_{\ell+2}} g_{BP,\ell}\left(u - \tilde{s}_\ell, \tilde{d}_\ell\right) du = \tilde{y}_\ell$$

from which a matrix equation $\tilde{\mathbf{G}}\tilde{\mathbf{c}} = \tilde{\mathbf{q}}$ can be established,

$$\tilde{\mathbf{q}} = [\tilde{y}_\ell]_{\ell \in \mathbb{Z}} \text{ and}$$

$$\tilde{\mathbf{G}} = [\tilde{G}_{\ell k}]_{\ell,k \in \mathbb{Z}} = \left[\int_{\tilde{t}_\ell}^{\tilde{t}_{\ell+1}} g_{BP,k}\left(u - \tilde{s}_k, \tilde{d}_k\right) du\right]_{\ell,k \in \mathbb{Z}}$$

Then the coefficients $\{\tilde{c}_\ell\}$ can be estimated as $\tilde{\mathbf{c}} = [\tilde{c}_\ell]_{\ell \in \mathbb{Z}} = \tilde{\mathbf{G}}^+ \tilde{\mathbf{q}}$ where $\tilde{\mathbf{G}}^+$ is the pseudo-inverse of $\tilde{\mathbf{G}}$.

## 4. SIMULATIONS

Illustrative simulations are conducted for an amplitude-and-phase modulated bandpass signal

$$x(t) = 2\sin(\omega_1 t)/(\omega_1 t)\cos(\omega_0 t + \sin(\omega_2 t)/(\omega_2 t))$$

with prameters $\omega_1 = 2\pi \times 10\,\text{rad/sec}$, $\omega_2 = 2\pi \times 2.5\,\text{rad/sec}$ and $\omega_0 = 2\pi \times 50\,\text{rad/sec}$. It is seen from Fig. 4 that the signal is a bandpass signal with center frequency 50Hz and bandwidth 30Hz, and is bounded by the amplitude 2. The bandwidth is chosen to be slightly larger than -3dB bandwidth and is convenient for the setting of TEM parameters. Both single-channel TEM and two-channel TEM are simulated. For the single-channel case, the signal is taken as a bandlimited one with upper-edge radian frequency $2\pi \times 65\,\text{rad/sec}$. Letting $T_{TEM}^{LP} = 1/(2 \times 65)\,s$, $\kappa = 1$ and $b = 1 + c$, the threshold can be set to be $\delta = T_{TEM}^{LP}/2$. Similarly, for two-channel TEM, letting $T_{TEM}^{BP} = 1/30\,s$, $\kappa = 1$ and $b = 1 + c$, the threshold can be set to be $\delta = T_{TEM}^{BP}/2$. In addition, $\omega_l = 2\pi \times 35\,\text{rad/sec}$ is used to calculate the interpolation function (9).

The sampling and reconstruction are shown in Fig. 5 and Fig. 6, respectively. It is seen that both one-channel and two-channel TEMs well reconstruct the input waveform. However, by comparing their spiking times, as expected, the two-channel TEM allows large TEM interval.

## 5. CONCLUSIONS

In this paper, we consider a sampling and reconstruction framework for bandpass signals by using multi-channel IF-TEMs. The similarities between PNS and TEM are exploited and a reconstruction approach is proposed. Compared to single-channel TEM for bandpasss signals, the multi-channel scheme allows large TEM interval and hence is convenient for implementations. Simulations demonstrate its feasibility and performance.